\documentclass[twocolumn]{revtex4}
\usepackage{amsmath}
\usepackage{amssymb}
\usepackage{amsthm}
\usepackage{amsfonts}
\usepackage{algorithmic}
\usepackage{enumerate}
\usepackage{latexsym}
\usepackage[dvips]{graphicx}

\begin{document}

\title{Coherent transport of armchair graphene constrictions}
\author{HuiQiong Yin$^{1,2}$}
\author{Wei Li$^{1}$}
\author{Xiao Hu$^{2}$}
\author{Ruibao Tao$^{1}$}
\affiliation{$^{1}$ Department of Physics, Fudan University, Shanghai 200433, China}
\affiliation{$^{2}$WPI Center for Materials Nanoarchitectonics, National Institute for
Materials Science, Tsukuba 305-0044, Japan}
\date{\today }

\begin{abstract}
The coherent transport properties of armchair graphene
nanoconstrictions(GNC) are studied using tight-binding approach and
Green's function method. We find a non-bonding state at zero Fermi
energy which results in a zero conductance valley, when a single
vacancy locates at $y=3n\pm1$ of a perfect metallic armchair
graphene nanoribbon(aGNR). However, the non-bonding state doesn't
exist when a vacancy locates at y=3n, and the conductance behavior
of lowest conducting channel will not be affected by the vacancy.
For the square-shaped armchair GNC consisting of three metallic aGNR
segments, resonant tunneling behavior is observed in the single
channel energy region. We find that the presence of localized edge
state locating at the zigzag boundary can affect the resonant
tunneling severely. A simplified one dimensional model is put
forward at last, which explains the resonant tunneling behavior of
armchair GNC very well.
\end{abstract}

\maketitle

\section{\label{sec1} Introduction}

Owing to many good properties, such as the ultra high Fermi velocity($%
10^{6}m/s$), the stability due to the sp2 hybridization and the feasibility
of large-scale integration, graphene has been regarded as a promising
candidate material for post-silicon electronics, and has stirred intensive
studies\cite%
{Novoselov,Wallace,Geim1,Kim1,Castro,Katsnelson,Fujita,Nakada,Wakabayashi,Peres,Brey,T.C.Li}%
. Many graphene microstructures have been studied both theoretically and
experimentally for future applications, such as the field-effect
transistors, the p-n junctions, metallic-metallic(semiconducting)
nanojunctions, L-shaped, Z-shaped, T-shaped and cross shaped junctions\cite%
{Obradovic,Q.Yan,Liang,Williams,Abanin,SQF,J.Li,S.Hong,Yisong,Y.P.Chen,Z.F.Wang,Z.P.Xu,Thushari,JYY,Yanyang,Areshkin}%
.

Graphene nanoconstrictions(GNC) is one of the important building blocks for
the carbon based electronic circuits\cite%
{Rojas,A.Rycerz,Barbaros,Chuvilin,Darancet,Andriotis,Todd,Molitor,Rojas1}.
With the combination of etching and deposition techniques, a
complete turn off of electrical transport of GNC has been
demonstrated as a function of the local gate voltage
experimentally\cite{Barbaros}. Furthermore, stable and rigid carbon
atomic chains were experimentally realized by removing carbon atoms
row by row from graphene, which shows the sophisticated technique to
control the shape of graphene nanodevices\cite{Chuanhong}.
Theoretical researches mainly focus on two types of GNCs, i.e., the
type based on zigzag graphene nanoribbon(zGNR), and the other type
based on armchair graphene nanoribbon(aGNR). The wedge-shaped GNCs
based on zGNR show a gap in the transmission spectrums, and the zero
conductance is related to the appearance of localized zero energy
edge states\cite{Rojas}. Based on a zigzag edged GNC, valley filter
is proposed which can produce a valley polarized current. Two valley
filters in series may function as an electrostatically controlled
valley valve\cite{A.Rycerz}. As for the GNCs based on aGNR,
performance limits of GNR field-effect transistors are investigated
by studying a GNC with two semiconducting wide aGNRs attached to a
narrow semiconducting one\cite{Rojas1}.

In this paper, we first discuss the lattice vacancy effects on the transport
properties of metallic aGNR. We then study the coherent transport properties
of GNCs with two wide metallic aGNRs attached to a narrow metallic one with
two vertical zigzag boundaries(ZBs). We investigate the role of the ZBs of
GNCs in the coherent transport by breaking the ZBs one by one. Using a one
dimensional model for the GNCs, we are able to understand the resonant
tunneling behaviors of the GNCs. Throughout the paper, we use the nearest
tight-binding approach and the recursive Green's function method to analyze
the transmission rate and the local density of states(LDOS)\cite%
{Sancho,Leonor,Datta,T.C.Li}.

The paper is organized as follows. In Sec.~\ref{sec2} we discuss the lattice
vacancy effects on the transport properties of metallic aGNRs. In Sec.~\ref%
{sec3}, the transport properties of armchair GNC are studied, and the
simplified model is put forward. In Sec.~\ref{sec4}, the summary is given.

\section{\label{sec2}Single lattice vacancy effect on metallic aGNR}

In the nearest tight-binding model, a perfect zGNR is always metallic
because of the finite overlap of the two edge states\cite{Wakabayashi1}.
However, aGNR can be either metallic (for $N=3n-1$) or semiconducting
(otherwise), where $N$ is the width of of aGNR counted by the number of
dimmer lines in the transverse direction with $n$ an integer\cite{Peres}. It
has been shown that a single vacancy at the edge of aGNR would induce a zero
conductance dip in the conductance spectrum\cite{T.C.Li}. It has also been
found that an on-site defect at different positions of aGNR induce different
conductance spectrums\cite{JYY}. It is then interesting to investigate the
transport property in the presence of a single vacancy at various positions
of aGNR.

 The tight-binding Hamitonian of the aGNR is given by
\begin{equation}
H_{aGNR}=\sum_{i}\varepsilon _{i}c_{i}^{\dag }c_{i}+t\sum_{\langle
i,j\rangle}c_{j}^{\dag }c_{i}
\end{equation}
where $\varepsilon _{i}$ is the on-site energy of atom $i$, $c_{i}^{\dag }$($%
c_{i}$) the creation(annihilation) operator at site $i$, and $t$ the
hopping constant between the nearest neighbors.

\begin{figure}[h]
\includegraphics[width=8.5cm]{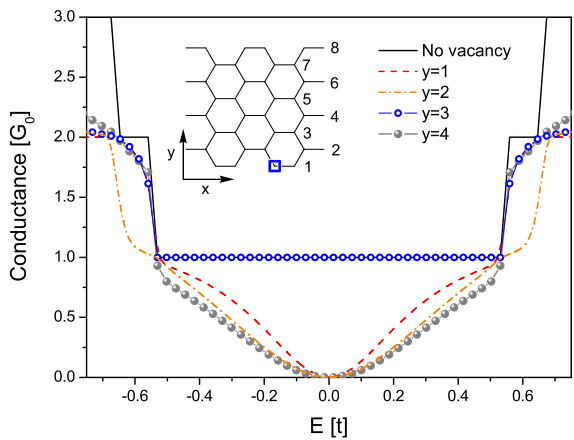}
\caption{(Color online). Conductance spectrums for a perfect
aGNR(solid line) with width 8, and for the case that a single
vacancy locates at y=1(dashed line), y=2(dash dotted line),
y=3(solid line with open circle), and y=4(solid line with solid
sphere). The conductance is in units of $G_{0}=2e^{2}/h$, where $h$
is the Plank constant.} \label{g1}
\end{figure}

 A lattice vacancy can be described by a very large on-site
energy in the tight-binding model \cite{L.Chico,Wakabayashi2}. Due
to the translational invariance of a perfect aGNR in the $x$ axis,
we only have to consider the effect of position of the vacancy
along the $y$ axis. Here we choose the width $N=8$ as an example, see Fig.~%
\ref{g1}. The vacancy is marked by a square. With the consideration
of symmetry along the $y$ axis, we only have to consider the vacancy
at the four positions, i.e. , $y=1$, $2$, $3$ and $4$.

The conductance spectrums are shown in Fig.~\ref{g1}. The solid line is the
conductance spectrum for a perfect aGNR, which shows quantized conductance
plateaus. The dashed line, dash dotted line, and solid line with solid
spheres represent the cases that a single vacancy resides at $y=1$, $y=2$,
and $y=4$. All the three lines show a conductance valley with a zero
conductance at $E=0$. The solid line with open circles shows the conductance
spectrum when a single vacancy locates at $y=3$. We see that the conductance
spectrum maintains perfectly the first quantized conduction plateau. The
presence of the vacancy doesn't affect the lowest conducting channel of
metallic aGNR at all.

\begin{figure}[h]
\includegraphics[scale=0.7]{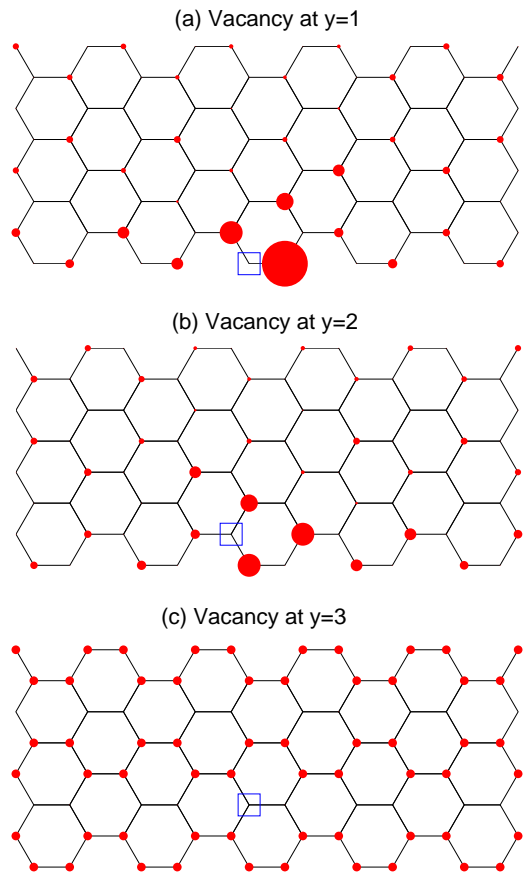}
\caption{(Color online). The LDS distribution of perfect aGNR (a), and the
aGNR with a vacancy at $y=1,E=0t$ (b), $y=2,E=0t$ (c), $y=3,E=0t$ (d), with
the width N=8}
\label{g2}
\end{figure}

To understand the peculiar effects of single lattice vacancy on the
metallic aGNR, we examine the microscopic distribution of LDOS of
the ribbon. Figure~\ref{g2}(a) shows the LDOS distribution at $E=0$
when a single vacancy locates at $y=1$, as marked by the square,
where the magnitude of the LDOS is denoted by the radius of the
solid circle. We can see that for the sublattice to which the
vacancy belongs(say A sublattice), all the atoms have zero LDOS.
Finite LDOS appear only on the other sublattice(B sublattice). It is
this non-bonding state that gives rise to the zero conductance at
$E=0$, as shown in Fig.~\ref{g1}. When the energy deviates from
zero, LDOS on A sublattice becomes finite, which opens possible
hopping paths for electrons to cross the vacancy region, and thus
provides finite conductance. Figure~\ref{g2}(b) shows the LDOS
distribution when the vacancy locates at $y=2$. We can see that the
non-bonding state also appears. When the vacancy locates at $y=4$,
the situation is similar with that of $y=1$, and $y=2$. However, for
a vacancy at $y=3$, we see in Fig.~\ref{g2}(c) that the LDOS
distribution is uniform at $y=3n\pm 1$, and zero at $y=3n$. The
pattern is the same throughout the energy region of the first
conductance channel, which explains why the lowest conductance
plateau is unaffected by the presence of a vacancy at $y=3$ shown in
Fig.~\ref{g1}.

We first notice that the response of the system to a vacancy at $y=3n$ can
be explained by the wave function of a perfect aGNR \cite{HHLin,Brey,ZHX},
\begin{equation}
\Phi _{k}\left( y\right) =\frac{1}{\sqrt{N+1}}\left(
\begin{array}{c}
\sin \left( \frac{2\pi }{3}y\right) \\
se^{i\theta (\mathbf{k})}\sin \left( \frac{2\pi }{3}y\right)%
\end{array}%
\right),  \label{eq10}
\end{equation}%
with $s=\pm 1$ for electrons and holes, $\theta (\mathbf{k})$ the phase
difference between two sublattices. The wave function exhibits nodes at $y=3n
$, which makes these sites insensitive to introduction of vacancy.

In order to understand the phenomenon that a single vacancy reduces
the LDOS at all the atoms of the same sublattice to zero at $E=0$,
we go back to the Shr\"odinger equation for the tight-binding model.
The wave functions at the three
nearest neighbors of the atom $i$ should satisfy the relation%
\begin{equation}
t\left( \phi _{1}^{B}+\phi _{2}^{B}+\phi _{3}^{B}\right)=E\phi _{i}^{A}.
\end{equation}
At zero energy, this implies
\begin{equation}
\phi _{1}^{B}+\phi _{2}^{B}+\phi _{3}^{B}=0.  \label{eq10a1}
\end{equation}
\begin{figure}[h]
\includegraphics[scale=0.6]{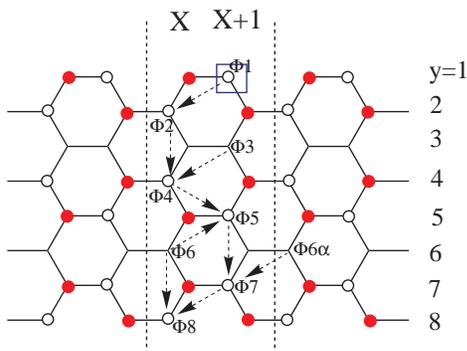}
\caption{(Color online). A single vacancy at the edge of the aGNR. }
\label{g3}
\end{figure}
Without lost of generality, we take aGNR with width $N=8$ with a
vacancy on the B sublattice, as shown by Fig.~\ref{g3}. Now we set a
vacancy at the edge as marked by the square in Fig.~\ref{g3}, which
renders $\Phi_1=0$. From the Shr\"odinger equation we can
immediately get $\Phi_{2}=0$. Then from $\Phi_{3}=0$, due to the
node of wave function of aGNR, and $\Phi_{2}=0$, we can deduce
$\Phi_{4}=0$. In a similar way we can show that all the atoms of
sublattice B have zero LDOS. It is easy to
see that the same discussion applies for the vacancy at any positions of $%
y=3n\pm 1$ of the aGNR.

\begin{figure}[h]
\includegraphics[width=4.5cm]{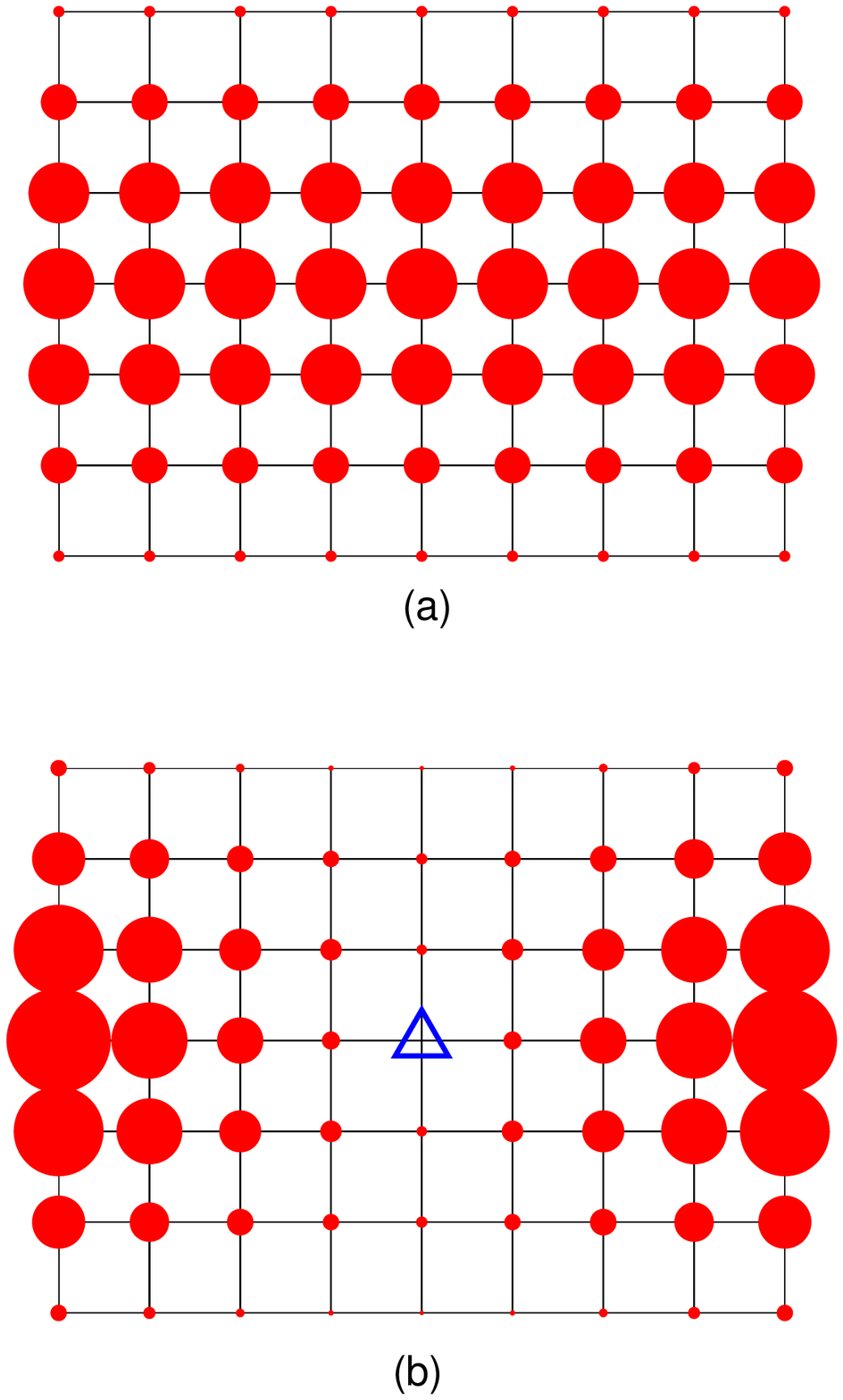}
\caption{(Color online). LDOS distribution of square lattice ribbon without
(a) and (b) with a single vacancy. The triangle denotes the vacancy. }
\label{g3a}
\end{figure}

The appearance of such non-bonding state is closely related with the
topological structure of graphene. A single vacancy cannot induce such
non-bonding state for the square lattice, as shown by Fig.~\ref{g3a}, where
the triangle denotes the vacancy.

\section{\label{sec3} Transport properties of armchair GNC}

\begin{figure}[h]
\includegraphics[width=8.0cm]{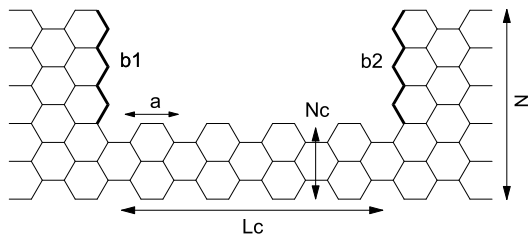}
\caption{Schematic diagram of the square shaped GNC. $N=11$ and $N_{\mathrm{c%
}}=5$ denote the widths of the wide aGNRs and narrow aGNR respectively. $L_{%
\mathrm{c}}=4$ is the length of central narrow aGNR in units of $a$. }
\label{g4}
\end{figure}

We now turn our attention to the armchair GNC shown in Fig.~\ref{g4}, with
widths of the wide metallic aGNRs and the narrow one denoted by $N$ and $N_{%
\mathrm{c}}$, and the length of the central segment $L_{\mathrm{c}}$ in
units $a$. Two vertical aligned zigzag boundaries are emphasized by the bold
lines with the notation $b1$, $b2$. We first study the conductance as a
function of $L_{\mathrm{c}}$.
\begin{figure}[h]
\includegraphics[width=8.5cm]{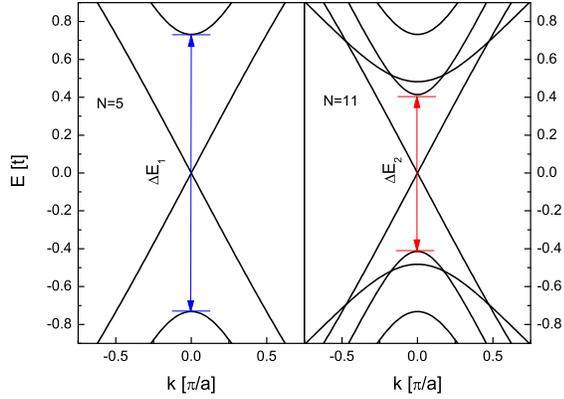}
\caption{(Color online). Band structures of perfect aGNRs with $N=5$ and $%
N=11$. $\Delta E_{1}$ and $\Delta E_{2}$ denote the single channel regions.}
\label{g5}
\end{figure}
\begin{figure}[h]
\includegraphics[width=8.5cm]{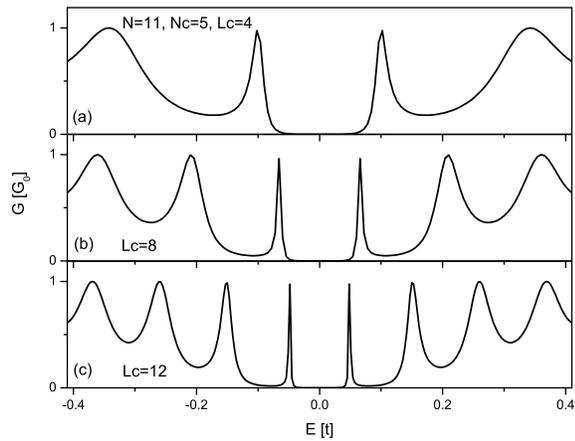}
\caption{(Color online). Conductance spectrums of square shaped GNC for $N=11
$ and $N_{\mathrm{c}}=5$, and $L_{\mathrm{c}}$ equals to 4(a), 8(b), and
12(c). }
\label{g6}
\end{figure}
The band structures of aGNR with width 5 and 11 are shown in
Fig.~\ref{g5}. In this paper, we will focus on the energy regime
$\Delta E_{2}$ around the zero energy, in which both the wide and
narrow aGNRs exhibit the single conduction channel. As in
Fig.~\ref{g6} we observe oscillating conductance with the peak value
reaching $G_{0}$, and a larger $L_{\mathrm{c}}$ induces more rapid
oscillations. Around $E=0$, the conduction is suppressed to zero for
all $L_{\mathrm{c}}>1$
\begin{figure}[h]
\includegraphics[width=8.5cm]{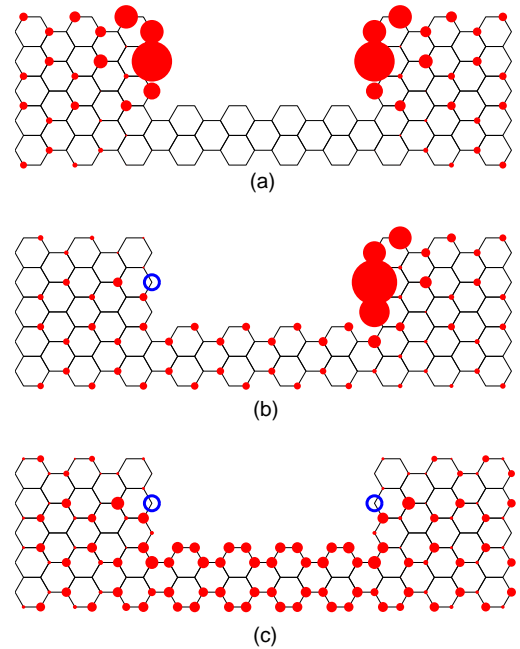}
\caption{(Color online). LDOS distribution for a GNC with two ZBs (a), one
ZB (b), and no ZB (c). The radius of the solid circle(red) stands for the
magnitude of LDS. The blank ring(blue) in panel (b) and (c) means the
vacancy. $N=11$, $Nc=5$, $Lc=4$. }
\label{g7}
\end{figure}
\begin{figure}[h]
\includegraphics[width=8.5cm]{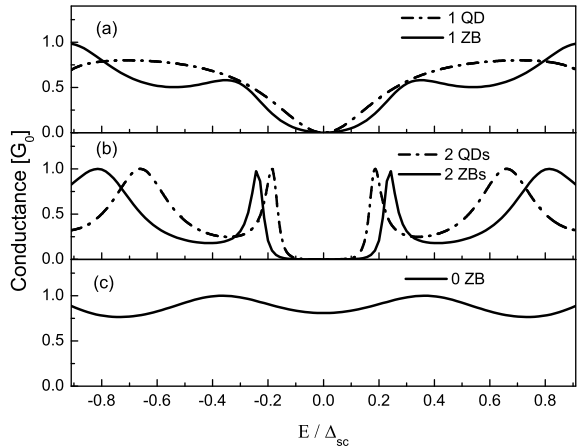}
\caption{(Color online). Conductance spectrums for GNCs with (a) one
ZB, (b) two ZBs, and (c) no ZB. The dashed curves are for the
nanowire and QD model. Energy is normalized by $\Delta{\mathrm sc}$,
the half-width of the single channel region ($\Delta E_{2}$ for GNC
and 2t for QD model).} \label{g9}
\end{figure}
\begin{figure}[h]
\includegraphics[width=8.0cm]{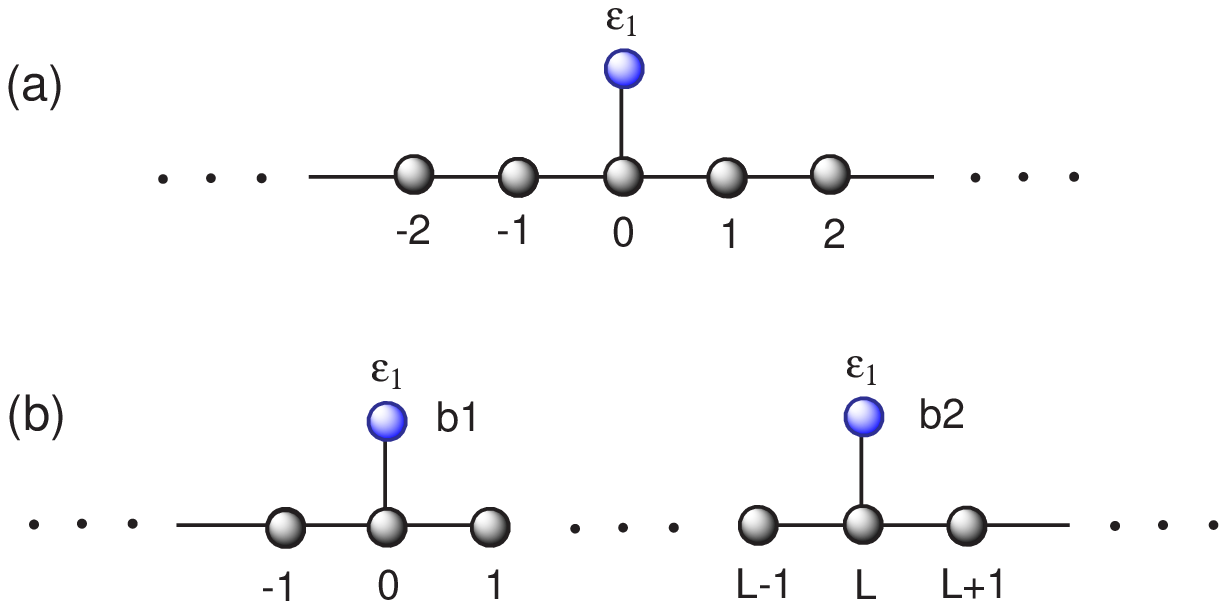}
\caption{(Color online). Schematic diagram of the model in which a quantum
wire couples to a single QD (a) and two QDs (b). $\protect\varepsilon _{1}$
is the eigenenergy of the QD(s).}
\label{g8}
\end{figure}

To understand the peculiar behaviors of the conductance of the GNC,
we study the LDOS distribution. As in Fig.~\ref{g7}(a), there is a
localized edge state at each ZB. We have also investigated ZB's with
different heights  (4, 5, 6, 7, 8, 9 for GNC with wider aGNR $N=23$)
and found that a ZB with three successive sites in a sequence is
enough to induce localized edge state, consistent with previous
works \cite{Nakada,Yisong,O.Hod}. The localized edge state is
suppressed by
the introduction of a single vacancy at the ZB b1 and b2, as shown in Figs.~%
\ref{g7}(b) and (c).

Next we study how the localized edge states affect the transport behaviors
by comparing the conductance spectrums of the three GNCs shown in Fig.~\ref%
{g7}. As shown in Figs.~\ref{g9} (a) and (b), for the GNCs with one
or two ZBs, the conductance is suppressed to zero near the zero
Fermi energy. However, for a GNC without ZB, the conductance remains
finite at $E=0$, see Fig.~\ref{g9} (c). Therefore, the localized
state at the ZB is responsible for the suppression of conductance to
zero at $E=0$.

Since we are considering the energy region where both the wide aGNRs and the
narrow one permit only one conducting channel, they can be modeled by a
single nanowire. A ZB which induces localized state can be modeled by a
quantum dot (QD) coupled to the nanowire. The model system is shown in Fig.~%
\ref{g8}. Here we would like to discuss the simpler case first, i.e., the
model with a single QD. For this system, Orellana \textit{et al.}\cite%
{Orellana} have already solved the transmission coefficient by setting the
wave function as
\begin{equation}
\psi =\left\{
\begin{array}{ccc}
e^{ikx}+re^{-ikx} & , & x<0 \\
te^{ikx} & , & x>0%
\end{array}%
\right.  \label{eq10a}
\end{equation}%
and matching the wave functions from both the left and right sides to that
coupled by the QD. The conductance can be derived analytically as
\begin{equation}
G=\frac{2e^{2}}{h}T=\frac{2e^{2}}{h}\frac{\left( \varepsilon -\varepsilon _{1}\right) ^{2}}{%
\left( \varepsilon -\varepsilon _{1}\right) ^{2}+\left( \frac{v}{2\sin ka}%
\right) ^{2}}  \label{eq11}
\end{equation}%
where $\varepsilon =2v\cos ka$ is the Fermi energy of the quantum wire, $%
\varepsilon_{1}$\ is the eigen energy of the side coupled by QD, $v$
is the hopping constant both within the nanowire and between the
nanowire and QD. From Eq. (\ref{eq11}) one can find that an
antiresonance appears at the energy $\varepsilon =\varepsilon _{1}$.
The localized edge state at zero Fermi energy\cite{Fujita} implies
$\varepsilon _{1}=0$, and explains why a single ZB of GNC induces a
zero conductance at the zero Fermi energy. Figure.~\ref{g9}(a) shows
the transmission spectrum of GNC with one ZB, which is well
described by the simple model especially in the small $k$ region.

Similarly, the GNC with two ZBs, as shown in Fig.~\ref{g7} (a), can
be modeled by two QDs coupled to a nanowire, as shown by
Fig.~\ref{g8} (b). Now we extend the approach by Orellana \textit{et
al.}\cite{Orellana} to the model with two QDs. The wave function
should be
\begin{equation}
\psi =\left\{
\begin{array}{ccc}
e^{ikx}+re^{-ikx} & , & x<0 \\
t^{\prime }e^{ikx}+r^{\prime }e^{-ikx} & , & 0<x<L \\
te^{ik\left( x-L\right) } & , & L<x.%
\end{array}%
\right.  \label{eq12}
\end{equation}%
Solving the wave function matching equations, we obtain the transmission
coefficients
\begin{equation}
t=\frac{2i\frac{\sin ^{2}k}{\sin kL}}{\left( \frac{\sin k}{\sin kL}\right)
^{2}-\left( e^{ik}-c+\frac{\sin k\left( L-1\right) }{\sin kL}\right) ^{2}}
\label{eq13}
\end{equation}%
where $c=\frac{1}{v}\left( \varepsilon -\frac{1}{\varepsilon -\varepsilon
_{1}}\right) $. Since $c$ diverges at $\varepsilon =\varepsilon _{1}=0$, one
has $t=0$, and thus the transmission probability $T=tt^{*}=0$. Figure~\ref%
{g9}(b) shows the transmission spectrum of the model of two QDs. We
find that the model can describe the GNC with two ZBs very well.

\begin{figure}[h]
\includegraphics[width=8.0cm]{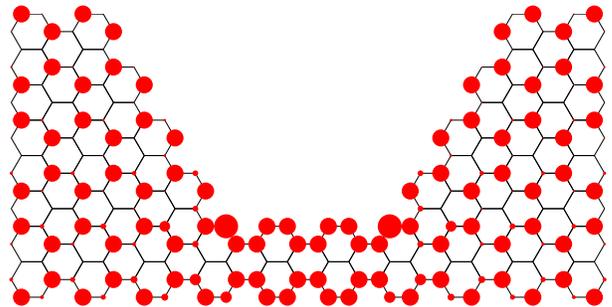}
\caption{(Color online). LDOS of GNC with armchair boundaries, with
 $N=17$, $N_{\mathrm c}=5$ and $L_{\mathrm c}=4$.} \label{g10}
\end{figure}

In the absence of ZB in the GNCs, there will be no localized state,
and thus no antiresonance, which results in a normal oscillation of
conductance without zero conductance valley, as shown by
Fig.~\ref{g9} (c). We also investigated GNCs with armchair
boundaries, and found no localized edge states, as shown in
Fig.~\ref{g10}, associated with finite conductance.

\section{\label{sec4} Summary}

Using tight-binding approach and Green's function method, we show
that for a metallic aGNR with a single vacancy at $y=3n\pm1$, the
zero conductance valley arises from a non-bonding state at the zero
Fermi energy. Next, for the square-shaped armchair GNC consisting of
three metallic aGNR segments, we find resonant tunneling behaviors
in the energy regime with single conduction channel. It is shown
that the localized edge state at the ZB affects the transport
properties severely. A one dimensional model with one or two QDs
coupled to a nanowire is put forward which can explain the resonant
tunneling behavior of aGNC very well.

\begin{acknowledgments}
The authors would like to thank P. A. Orellana and A. Tanaka for helpful
discussions.
\end{acknowledgments}


\begin{thebibliography}{99}
\bibitem{Novoselov} K. S. Novoselov, A. K. Geim, S. V. Morozov, D. Jiang, Y.
Zhang, S. V. Dubonos, I. V. Grigorieva, and A. A. Firsov, Science \textbf{306%
}, 666 (2004).

\bibitem{Wallace} P. R. Wallace, Physical Review \textbf{71}, 622 (1947).

\bibitem{Geim1} A. K. Geim and K. S. Novoselov, Nat Mater \textbf{6} ,
183(2007).

\bibitem{Kim1} M. S. Purewal, Y. Zhang and P. Kim, Phys. Stat. Sol.(b)
\textbf{243}, 3418 (2006).

\bibitem{Castro} A. H. Castro Neto, F. Guinea and N. M. R. Peres, K. S.
Novosolev and A. K. Geim, Rev. Mod. Phys. {\textbf{81}}, 109 (2009).

\bibitem{Katsnelson} M. I.Katsnelson, Materials Today \textbf{10}, 20(2007).

\bibitem{Fujita} M. Fujita, K. Wakabayashi, K. Nakada and K. Kusakabe,J.
Phys. Soc. Jpn. \textbf{65} 1920,(1996)

\bibitem{Nakada} K. Nakada, M. Fujita, G. Dresselhaus, and M. S.
Dresselhaus, Phys. Rev. B. \textbf{54}, 17954 (1996).

\bibitem{Wakabayashi} K. Wakabayashi, Physical Review B \textbf{64}, 125428
(2001).

\bibitem{Peres} N. M. R. Peres, A. H. Castro Neto, and F. Guinea, Phys. Rev.
B \textbf{73}, 195411 (2006).

\bibitem{Brey} L. Brey and H. A. Fertig, Phys. Rev. B \textbf{73}, 235411
(2006)

\bibitem{T.C.Li} T. C. Li and S.-P. Lu, Phys. Rev.B \textbf{77}, 085408
(2008).

\bibitem{Obradovic} B. Obradovic, R. Kotlyar, F. Heinz, P. Matagne, T.
Rakshit, M. D. Giles, M. A. Stettlera, and D. E. Nikonov, Appl. Phys. Lett.
\textbf{88}, 142102 (2006)

\bibitem{Q.Yan} Q. Yan, B. Huang, J. Yu, F. Zheng, J. Zang, J. Wu, B.-L. Gu,
F. Liu, and W. Duan, Nano Lett. \textbf{7}, 1469(2007)

\bibitem{Liang} G. Liang, N. Neophytou, M. S. Lundstrom and D. E. Nikonov,
J. Appl. Phys. \textbf{102}, 054307 (2007).

\bibitem{Williams} J. R. Williams, L. DiCarlo, C. M. Marcus, Science \textbf{%
317}, 638 (2007)

\bibitem{Abanin} D. A. Abanin and L. S. Levitov, Science \textbf{317}, 641
(2007)

\bibitem{SQF} W. Long, Q. F. Sun, and J. Wang, Phys. Rev. Lett. \textbf{101}%
, 166806 (2008).

\bibitem{J.Li} J. Li, S. Q. Shen, Phys. Rev. B \textbf{78}, 205308 (2008).

\bibitem{S.Hong} S. Hong, Y. Yoon, and J. Guo, Appl. Phys. Lett. \textbf{92}%
, 083107 (2008).

\bibitem{Yisong} H. Li, L. Wang and Y. Zheng, cond-mat/0808.0947

\bibitem{Y.P.Chen} Y. P. Chen, Y. E. Xie, and X. H.Yan,, J. Appl. Phys.
\textbf{103}, 063711(2008).

\bibitem{Z.F.Wang} Z. F. Wang, Q. W. Shi, Q. Li, X.Wang, J.G. Hou, H. Zheng,
Y. Yao, J. Chen, Appl. Phys. Lett. \textbf{91}, 053109 (2007).

\bibitem{Z.P.Xu} Z. P. Xu and Q. S. Zheng, Appl. Phys. Lett. \textbf{90},
223115 (2007).

\bibitem{Thushari} T. Jayasekera and J. W. Mintmire, Nanotechnology \textbf{%
18}, 424033 (2007).

\bibitem{JYY} J.-Y. Yan, P. Zhang, B. Sun, H.-Z. Lu, Z. Wang, S. Duan, and
X.-G. Zhao, Phys. Rev. B \textbf{79}, 115403(2009).

\bibitem{Yanyang} Y. Zhang, J.-P. Hu, B. A. Bernevig, X. R. Wang, X. C. Xie,
and W. M. Liu, Phys. Rev. B \textbf{78}, 115413(2008).

\bibitem{Areshkin} D. A. Areshkin and C. T. White, Nano Lett. \textbf{7}, 3253 (2007)

\bibitem{Rojas} F. Mu\~{n}oz-Rojas, D. Jacob, J. Fern\'{a}ndez-Rossier, and
J. J. Palacios, Phys. Rev. B. \textbf{74}, 195417 (2006)



\bibitem{A.Rycerz} A. Rycerz, J. Tworzyd{\l }o, and C. W. J. Beenakker,
Nature Phys. \textbf{3}, 172(2007).

\bibitem{Barbaros} B. \"{O}zyilmaz, P. Jarillo-Herrero, D. Efetov, and P.
Kim, Appl. Phys. Lett. \textbf{91}, 192107 (2007).

\bibitem{Chuvilin} A. Chuvilin, J. C. Meyer, G. Algara-Siller, and U.
Kaiser, New J. Phys. \textbf{11},083019 (2009).

\bibitem{Darancet} P. Darancet, V. Olevano, and D. Mayou, Phys. Rev. Lett.
\textbf{102}, 136803(2009).

\bibitem{Andriotis} A. N. Andriotis, E. Richter, and M. Menon, Appl. Phys.
Lett. \textbf{91}, 152105 (2007).

\bibitem{Todd} K. Todd, H.-T. Chou, S. Amasha, and D. Goldhaber-Gordon, Nano
Lett. \textbf{9}, 416 (2009).

\bibitem{Molitor} F. Molitor, A. Jacobsen, C. Stampfer, J. G\"{u}ttinger, T.
Ihn, and K. Ensslin, Phys. Rev. B \textbf{79}, 075426(2009).

\bibitem{Rojas1} F. Mu\~{n}oz-Rojas, J. Fern\'{a}ndez-Rossier, L. Brey and
J. J. Palacios, Phys. Rev. B. \textbf{77}, 045301 (2008)

\bibitem{Chuanhong} Chuanhong Jin, Haiping Lan, Lianmao Peng, Kazu Suenaga,
and Sumio Iijima, Phys. Rev. Lett. \textbf{102}, 205501 (2009)

\bibitem{O.Hod} O. Hod, V. Barone, and G. E. Scuseria, Phys. Rev. B \textbf{%
77}, 035411 (2008).

\bibitem{Lopez} M. P. L. Sancho, J. M. L. Sancho, J. Rubio, J. Phys. F: Met.
Phys. \textbf{15}, 851 (1985).

\bibitem{L.Chico} L. Chico, V. H. Crespi, L. X. Benedict, S. G. Louie, and
M. L. Cohen, Phys. Rev. B, \textbf{54} 2600(1996).

\bibitem{Wakabayashi2} K. Wakabayashi, J. Phys. Soc. Jpn. \textbf{71}, 2500 (2002).

\bibitem{HHLin} H. H. Lin, T. Hikihara, H. T. Jeng, B. L. Huang, C .Y. Mou,
and X. Hu, Phys. Rev. B. \textbf{79}, 035405 (2009)

\bibitem{ZHX} H. Zheng, Z. F. Wang, Tao Luo, Q. W. Shi, and J. Chen, Phys.
Rev. B. \textbf{75}, 165414 (2007)

\bibitem{Orellana} P. A. Orellana, F. Dom\'{\i}nguez-Adame, I. G\'{e}mez,
and M. L. Ladr\'{o}n de Guevara, Phys. Rev. B. \textbf{67}, 085321 (2003)

\bibitem{Sancho} M. P. L. Sancho, J. M. L. Sancho, and J. Rubio, J. Phys. F:
Met. Phys. \textbf{14}, 1205 (1984).

\bibitem{Leonor} L. Chico, L. X. Benedict, S. G. Louie, and M. L. Cohen,
Phys. Rev. B. \textbf{54}, 2600 (1996)

\bibitem{Datta} S. Datta, \textit{Electronic Transport in Mesoscopic Systems}
(Cambridge University Press, Cambridge, 1995) .

\bibitem{Wakabayashi1} K. Wakabayashi, and M. Sigrist, Phys. Rev. Lett.
\textbf{84}, 3390 (2000).
\end{thebibliography}
\end{document}